\documentclass[conference]{IEEEtran}
\usepackage{graphicx}
\usepackage{times}
\usepackage{subfigure}
\usepackage{graphicx}
\usepackage{amsmath}
\usepackage{bm}
\usepackage{url}
\usepackage{stmaryrd}
\usepackage{balance}
\usepackage{amssymb}
\usepackage{pgfplots}
\usepackage{pifont}
\usepackage{epsfig}
\usepackage{booktabs}% professional tables
\usepackage{pgffor}
\usepackage{tikz}
\usepackage{multirow}
\usepackage{amsmath}

\makeatletter
\newcommand\xleftrightarrow[2][]{%
  \ext@arrow 9999{\longleftrightarrowfill@}{#1}{#2}}
\newcommand\longleftrightarrowfill@{%
  \arrowfill@\leftarrow\relbar\rightarrow}
\makeatother

\newcommand{\ie}[0]{\textit{i.e.}}

\newcommand{\mb}{\mathbf}

\newcommand{\our}{\textsc{LATTE}}

\newcommand{\autoencoder}{\textsc{Auto-Encoder}}

\newcommand{\deepwalk}{\textsc{DeepWalk}}
\newcommand{\nodevec}{\textsc{Node2Vec}}

\usepackage{algorithm}
\usepackage{algorithmic}

\begin{document}
%----------------------------------------------------------------------------------------
\title{LATTE: Application Oriented Social Network Embedding}
\author{
\IEEEauthorblockN{Lin Meng$^\star$, Jiyang Bai$^\star$, Jiawei Zhang$^\star$}
\IEEEauthorblockA{$^\star$IFM Lab, Department of Computer Science, Florida State University, FL, USA\\
			      lin@ifmlab.org, jiyang@ifmlab.org, jiawei@ifmlab.org}
}

%author{\IEEEauthorblockN{1\textsuperscript{st} Given Name Surname}
%	\IEEEauthorblockA{\textit{dept. name of organization (of Aff.)} \\
%		\textit{name of organization (of Aff.)}\\
%		City, Country \\
%		email address}
%	
\maketitle

%%
%\author{ Lin~Meng$^\star$, Jiyang~Bai$^\star$, Jiawei~Zhang$^\star$, Yanjie~Fu$^\dagger$ }
%\affiliation{%
%  $^\star$Department of Computer Science, Florida State University, FL, USA \\
%$^\dagger$Department of Computer Science, Missouri University of Science and Technology, MO, USA 
%}
%\email{lin@ifmlab.org, jiyang@ifmlab.org,   jiawei@ifmlab.org,  fuyan@mst.edu}
%

\begin{abstract}

In recent years, many research works propose to embed the network structured data into a low-dimensional feature space, where each node will be represented as a feature vector. However, due to the detachment of the embedding process with external tasks, the learned embedding results by most existing embedding models can be ineffective for application tasks with specific objectives, e.g., network alignment, community detection or information diffusion. In this paper, we propose to study the application oriented heterogeneous social network embedding problem. Significantly different from the existing works, besides the network structure preservation, the problem should also incorporate the objectives of external applications in the objective function. To resolve the problem, in this paper, we propose a novel network embedding framework, namely ``\textit{app\underline{L}ic\underline{A}tion orien\underline{T}ed ne\underline{T}work \underline{E}mbedding}'' ({\our}). In {\our}, the heterogeneous network structure can be applied to compute the node ``diffusive proximity'' scores, which capture both the local and global network structures. Based on these computed scores, {\our} learns the network representation feature vectors by extending the autoencoder model to the heterogeneous network scenario, which can also effectively unite the objectives of network embedding and external application tasks. Extensive experiments have been done on real-world heterogeneous social network datasets, and the experimental results have demonstrated the outstanding performance of {\our}.

\end{abstract}

%%\category{H.2.8}{Database Management}{Database Applications-Data Mining} 
%\keywords{Network Embedding; Aligned Heterogeneous Networks; Link Prediction; Community Detection; Data Mining}


%\maketitle
%-----------------------------------------------

\section{Introduction}\label{sec:introduction}
%In this paper, we will take online social networks as an example to illustrate the problem settings and proposed models.
In the era of big data, a rapidly increasing number of online websites emerge to provide various services, which can be represented as heterogeneous and complex networks. The representative examples include \textit{online social networks} (e.g., Facebook and Twitter), \textit{e-commerce sites} (e.g., Amazon and eBay) and \textit{academic sites} (e.g., DBLP and Google Scholar). These network data are hard to deal with due to their \textit{complex structures}, containing various types of nodes and links. In addition, great challenges exist in handling the complex network data with traditional machine learning algorithms, which usually take feature vectors as the input and cannot be applied to networked data directly.

In recent years, many research works propose to embed the social network data into a low-dimensional feature space \cite{BUGWY13,LLSLZ15,PAS14,TQWZYM15,GL16}, in which each node is represented as a feature vector. With these embedding feature vectors, the original network structure can be effectively reconstructed, classic learning algorithms can be applied directly, and the representations can also be widely used in external applications. Application tasks like \textit{network alignment}, \textit{community detection} and \textit{information diffusion} are extremely important for online social network studies. \textit{Network alignment} \cite{KZY13} aims at inferring the set of \textit{anchor links} connecting the shared users across social networks, which are usually subject to the \textit{one-to-one} cardinality constraint. \textit{Community detection} \cite{L07} aims at partitioning social network users into different clusters with the objective that aggregating close users and parting seperated users.  While  \textit{information diffusion} \cite{DR01,KKT03} aims at modeling the information diffusion process in online social networks, with the objective to learn both the personal topic preference of users and the activation relationships among users. To ensure the embedding results applicable to these tasks, incorporating the task-specific objectives in the embedding process is desired and necessary.
%In this paper, we will tackle these challenges by proposing a novel extensible heterogeneous social network embedding model, which can effectively incorporate the objectives of external tasks in the learning process. 
%``users with close connections will be placed in the same cluster, while those without connections will be divided into different clusters''
%However, applying these existing embedding methods on real-world social networks may encounter several challenges: (1) \textit{data perspective}, the \textit{heterogeneity} of real-world social network structure renders existing homogeneous-network oriented embedding models \cite{TQWZYM15,GL16,PAS14} failing to work, (2) \textit{structure preserving perspective}, many first-order proximity based embedding methods \cite{TQWZYM15,CHTQAH15} can hardly preserve the complex and extensively connected social network structure, and (3) \textit{task perspective}, the detachment of the embedding process \cite{PAS14,TQWZYM15,GL16,CHTQAH15,WCZ16} with external tasks makes the learned results ineffective for application tasks with specific objectives. 
In this paper, we will study the \textit{Application Oriented Network Embedding} problem. Besides preserving the network structure, the problem also aims at incorporating the application oriented objectives into the embedding optimization function, so as to guarantee the learned embedding results can be effectively applied in external application tasks. Various external applications can be incorporated in the network embedding learning, and in this paper, we will take ``\textit{network alignment}'', ``\textit{community detection}'' and ``\textit{information diffusion}'' as the examples to illustrate the problem setting and proposed framework. We claim that the ``\textit{network embedding results learned from `application oriented network embedding' will achieve better performance in these specific applications}'' than other general network embedding models \cite{BUGWY13,LLSLZ15,PAS14,TQWZYM15,GL16,TQWZYM15,CHTQAH15,WCZ16}.
%The problem studied in this paper is a novel research problem, and it has never been studied by this context so far. The studied problem is different from the other existing network embedding works in various aspects, including (1) \textit{network data}: different from the homogeneous networks studied in \cite{TQWZYM15, GL16, PAS14}, the social network data studied in {\problem} is heterogeneous in both the network structure and node attributes; (2) \textit{structure preservation}: instead of merely preserving the local network structure based on $1_{st}$-order proximity \cite{TQWZYM15, CHTQAH15}, {\problem} aims at maintaining the network structure with $n_{th}$-order proximity instead ($n \in \{1, 2, \cdots, \infty\}$); and (3) \textit{external application}: besides the network structure preservation, {\problem} will also incorporate the objectives of external application in the model learning, which has never been studied in existing network embedding problems \cite{PAS14, TQWZYM15, GL16, CHTQAH15, WCZ16}. 
However, solving the ``application oriented network embedding'' problem is not an easy task, which may suffer from several great challenges:
%\vspace{-10pt}
\begin{itemize}
	\item \textit{Heterogeneity of Network}: The networks studied in this paper are heterogeneous networks, involving very complex network structures. A new framework which can incorporate heterogeneous information into a unified analytics is required and necessary.
	
	\item \textit{Local and Global Network Structure Preservation}: Besides preserving the local network structure (i.e., node local neighborhood), the network embedding results should also preserve the global network structure (i.e., node global connection patterns). A new network proximity measure that can capture both local and global structures will be desired.
	
	\item \textit{External Application Incorporation}: Incorporating the external application tasks in the network embedding is the main objective in this paper. How to effectively incorporate the external application tasks in the network representation learning is still an open question by this context so far.
\end{itemize}

To resolve the above challenges, we propose a novel network embedding framework, namely ``\textit{app\underline{L}ic\underline{A}tion orien\underline{T}ed ne\underline{T}work \underline{E}mbedding}'' ({\our}). Based on the heterogeneous networks, {\our} computes the node closeness scores based on a new measure named ``\textit{diffusive proximity}'', which can be generalized to capture node $n_{th}$-order proximity ($n \in \{1, 2, \cdots, \infty\}$) effectively. {\our} learns the network representation feature vectors based on the ``\textit{Collective Autoencoder}'' model, which can effectively integrate the objective functions of both network embedding and external application tasks. We will apply {\our} on the network embedding tasks for three different application problems, including \textit{network alignment}, \textit{community detection} and \textit{information diffusion} respectively.

%The remaining parts of this paper are organized as follows. We will provide the terminology definitions and problem formulation in Section~\ref{sec:formulation}. Introduction to the proposed {\our} model will be provided in Section~\ref{sec:method}, whose learning performance will be evaluated in Section~\ref{sec:experiment}. Finally, we will introduce the related works in Section~\ref{sec:related_work} and conclude this paper in Section~\ref{sec:conclusion}. 

%-----------------------------------------------
\section{Terminology Definition and Problem Formulation} \label{sec:formulation}

In this section, we will introduce the definitions of several important concepts used in this paper, and provide the formulation of the studied problem.

\subsection{Notations}\label{subsec:notation}

In the following sections, we will use the lower case letters (e.g., $x$) to represent scalars, lower case bold letters (e.g., $\mb{x}$) to denote column vectors, bold-face upper case letters (e.g., $\mb{X}$) to denote matrices, and upper case calligraphic letters (e.g., $\mathcal{X}$) to denote sets. We use $\mb{X}(i,:)$ and $\mb{X}(:,j)$ to represent $i_{th}$ row and $j_{th}$ column of matrix $\mb{X}$ respectively. $X(i,j)$ or $X_{i,j}$ denotes the ($i_{th}$, $j_{th}$) of $\mb{X}$. We use $\mb{X}^\top$ and $\mb{x}^\top$ to represent the transpose of matrix $\mb{X}$ and vector $\mb{x}$. For vector $\mb{x}$, we represent its $L_p$-norm as $\left\| \mb{x} \right\|_p = (\sum_i |x_i|^p)^{\frac{1}{p}}$. The Frobenius norm of matrix $\mb{X}$ can be represented as $\left\| \mb{X} \right\|_F = (\sum_{i,j} X_{i,j}^2)^{\frac{1}{2}}$. The element-wise product of vectors $\mb{x}$ and $\mb{y}$ of the same dimension is represented as $\mb{x} \odot \mb{y}$, while the element-wise product of matrices $\mb{X}$ and $\mb{Y}$ is denoted as $\mb{X} \odot \mb{Y}$. Notation $\mb{Tr}(\mb{X})$ denotes the trace of matrix $\mb{X}$. 

\subsection{Terminology Definition}
We will study the \textit{heterogeneous social networks} in this paper, which includes a group of profile/textual content information for the nodes besides social connections. Formally, we can represent the studied heterogeneous social network as $G = (\mathcal{V}, \mathcal{E})$, involving the node and edge sets $\mathcal{V}$ and $\mathcal{E}$ respectively. To be more specific, we will take the Twitter and Foursquare social networks as examples to illustrate the network setting. In the Twitter network, for the user node, we can have their basic \textit{profile} information, including the \textit{user full name}, \textit{hometown}, etc. Meanwhile, for the post node in the studied network, their information covers \textit{textual content}, \textit{location checkins} and \textit{timestamps} contained in the posts. 
%\begin{definition}
%(Heterogeneous Social Networks): The \textit{heterogeneous social network} studied in this paper can be represented as a graph $G = (\mathcal{V}, \mathcal{E})$, where $\mathcal{V} = \mathcal{U} \cup \mathcal{P}$ denotes the set of users and posts, and $\mathcal{E} =\mathcal{E}_{u,u} \cup \mathcal{E}_{u,p}$ represents the set of social connections among users and write/like/reply/share links between users and posts. What's more, for the user nodes, we can also achieve their basic profile information; and for the post nodes, we can also achieve their textual contents and other diverse information.
%
%%for the nodes in $\mathcal{V}$, they can be associated with a set of attributes as well, which are denoted as set $\mathcal{T}$ = $\bigcup_i \mathcal{T}_i$. For each node $v \in \mathcal{V}$, we can represent its attributes as set $\bigcup_i \mathcal{T}_i(v)$, where $\mathcal{T}_i(v)$ represents the set of $i_{th}$-type attributes associated with $v$.
%\end{definition}
%
%Users nowadays are usually involved in multiple online social networks simultaneously, which can be effectively modeled with the \textit{aligned social network} concept and users in online social networks usually belong to different \textit{social communities}, and they tend to socialize more frequently with users within the same communities. Via the social interactions among users, information can propagate among users in social networks.
To study the application oriented network embedding problem, we first introduce several key definitions used in three tasks.

\noindent \textbf{Definition 1}
(Aligned Social Networks): Formally, the $n$ \textit{aligned social networks} can be represented as $\mathcal{G} = \big((G^{(1)}, \cdots, G^{(n)}),  (\mathcal{A}^{(1, 2)}, \cdots, \mathcal{A}^{(n-1,n)}) \big)$, where $G^{(1)}, \cdots, G^{(n)}$ denote the $n$ social networks and $\mathcal{A}^{(i,j)}, i,j \in \{1, 2, \cdots, n\}, i < j$ denotes the anchor link set between networks $G^{(i)}$ and $G^{(j)}$.
In this paper, we will take a pair of aligned networks, i.e., $\mathcal{G} = ((G^{(1)}, G^{(2)}), (\mathcal{A}^{(1, 2)}))$ as an example to illustrate the application problem settings.

\noindent \textbf{Definition 2}
	(Social Community): Given the user set $\mathcal{U}$ in a social network, its \textit{social community} structure can be represented as $\mathcal{C} = \{\mathcal{U}_1, \mathcal{U}_2, \cdots, \mathcal{U}_k\}$, where $\mathcal{U}_i \cap \mathcal{U}_j = \emptyset, \forall i, j \in \{1, 2, \cdots, k\} \land i \neq j$, and $\bigcup_{i=1}^k \mathcal{U}_i = \mathcal{U}$.

\noindent \textbf{Definition 3}
	(Infected Network): Formally, let $\mathcal{O}$ denote the set of topics propagated in the network. Given a certain topic $o_i \in \mathcal{O}$, its diffusion process in network $G$ can be represented as the \textit{infected network} $G_{o_i} = (\mathcal{U}_{o_i}, \mathcal{E}_{o_i})$ involving activated users $\mathcal{U}_{o_i} \subset \mathcal{U}$ and diffusion channels among users $\mathcal{E}_{o_i} \subset \mathcal{U} \times \mathcal{U}$. $\mathcal{E}_{o_i}$ is not necessarily a subset of link set $\mathcal{E}$ in network $G$, and $(u_i, u_j) \in \mathcal{E}_{o_i}$ iff $u_i$ activates $u_j$ via an indirect channel in the diffusion process of topic $o_i$.

\subsection{Problem Statement}
The \textit{application oriented social network embedding} problem studied in this paper aims at learning a mapping function $f: \mathcal{U} \to \mathbb{R}^d$ to project the user nodes in the network to a feature space of dimension $d$. Three objectives are covered in learning the mapping $f$:  (1) the if \textit{network alignment} is the oriented application task, structure between networks, i.e., the set of known anchor links in set $\mathcal{A}^{(1,2)}$ between networks $G^{(1)}$ and $G^{2)}$, (2) the social community structure of the network, i.e., $\mathcal{C} = \{\mathcal{U}_1, \mathcal{U}_2, \cdots, \mathcal{U}_k\}$, if \textit{community detection} is the oriented application task, and (3) the information diffusion process in the network, i.e., the infected networks $G_{o_i} = (\mathcal{U}_{o_i}, \mathcal{E}_{o_i}), \forall o_i \in \mathcal{O}$, if \textit{information diffusion} is the oriented application task.

%(1) effective fusion of heterogeneous network information, (2) local and global network structure preservation in the embedding results, and (3) incorporation of external application task objectives in model learning. To be more specific, the learned embedding feature vectors for the user nodes based on $f$ should be able to capture

%-----------------------------------------------
\section{Proposed Method}\label{sec:method}

In this paper, we will propose a novel network embedding model {\our} to learn the embedding feature vectors of nodes in online social networks, which can fuse the heterogeneous social network information in the learned feature representations and capture both local and global network structure. Furthermore, {\our} is also an easily extensible embedding model, where the objectives of external application tasks can be incorporated in the embedding process seamlessly. 

%-------------------------------------------------------------------------------------------------------------------------------------------------------------
%-------------------------------------------------------------------------------------------------------------------------------------------------------------
%-------------------------------------------------------------------------------------------------------------------------------------------------------------

\subsection{Heterogeneous Social Network Embedding Model}

Slightly different from the embedding problems of other types of data, like images or text, the nodes in networks are extensively connected, which will create extra constraints on the embedding feature vectors of the nodes learned from the raw input information, i.e., strongly connected nodes have similar representations. In this part, we will provide the descriptions of the raw features extracted from the heterogeneous network, and the isolated embedding model. The ``\textit{Collective Autoencoder}'' model and  the ``proximity constraints'' will be introduced in Section~\ref{subsec:embedding_model} in detail. 

%\subsubsection{Node Raw Feature Extraction}
Based on the diverse profiles, textual content, location check-ins, active timestamps information about the nodes in social networks, a set of raw features can be extracted for them in the social networks. Here, we will take the user node as an example to illustrate the raw feature extraction process. User's profile covers basic information about the user, including his/her name, gender, age and hometown. We propose to represent the profile information of user $u_i \in \mathcal{U}$ as a raw feature vector $\mb{x}^{p}_i = [\mb{x}^{p,n}_i, \mb{x}^{p,g}_i, {x}^{p,a}_i, \mb{x}^{p,h}_i]$. Except feature ${x}^{p,a}_i$, which is an integer indicating the user age, the remaining entries are all represented in a way similar to ``bag-of-word''. For instance, for the user's hometown information, the hometown locations of all users are listed first, and the entry ${x}^{p,h}_i(j)$ has value $1$ iff the $j_{th}$ location is $u_i$'s hometown, otherwise it will be $0$. It is similar to the feature vectors $\mb{x}^{p,n}_i, \mb{x}^{p,g}_i$ about user's name and gender information as well as the other types of nodes (e.g., the posts) in the networks, for which a set of raw features can be obtained as well. These extracted raw  features are usually of a large dimension, which will be fed into the model to be introduced in the following subsection to learn the embedding representations of users and posts respectively.
\subsubsection{Raw Feature Embedding with Autoencoder Model}\label{subsec:autoencoder}
Auto-encoder is an unsupervised neural network model, which projects the user/post nodes (from the original feature representations) into a low-dimensional feature space via a series of non-linear mappings. Auto-encoder model involves two steps: encoder and decoder. The encoder part projects the original feature vectors to the objective feature space, while the decoder step recovers the latent feature representation to a reconstruction space. In the auto-encoder model, we generally need to ensure that the original feature representations of user/post nodes should be close to the reconstructed feature representations.
Formally, let $\mb{x}_i$ represent the extracted feature vector for node $v_i \in \mathcal{V}$. Generally, feature vector $\mb{x}_i$ covers almost all the information we can obtain about the nodes, including ``who'' the user is, ``where'', ``when'' and ``what'' the users and posts are about. Via $o$ layers of projections, we can represent $\mb{y}^1_i, \mb{y}^2_i, \cdots, \mb{y}^o_i$ as the latent feature representation of the node at hidden layers $1, 2, \cdots, o$ in the encoder step, the encoding result in the objective feature space can be represented as $\mb{z}_i \in \mathbb{R}^{d}$ with dimension $d$. Formally, the relationship between these variables can be represented with the following equations:
$$\begin{cases}
\mb{y}^1_i &= \sigma (\mb{W}^1 \mb{x}_i + \mb{b}^1),\\
\mb{y}^k_i &= \sigma (\mb{W}^k \mb{y}^{k-1}_i + \mb{b}^k), \forall k \in \{2, 3, \cdots, o\},\\
\mb{z}_i &= \sigma (\mb{W}^{o+1} \mb{y}^o_i + \mb{b}^{o+1}).
\end{cases}$$

Meanwhile, in the decoder step, the input will be the latent feature vector $\mb{z}_i$ (i.e., the output of the encoder step), and the final output will be the reconstructed vector $\hat{\mb{x}}_i$. The latent feature vectors at each hidden layer can be represented as $\hat{\mb{y}}^{o}_i, \hat{\mb{y}}^{o-1}_i, \cdots, \hat{\mb{y}}^{1}_i$. The relationship among these vector variables can be denoted as
$$\begin{cases}
\hat{\mb{y}}^o_i &= \sigma (\hat{\mb{W}}^{o+1} \mb{z}_i + \hat{\mb{b}}^{o+1}),\\
\hat{\mb{y}}^{k-1}_i &= \sigma (\hat{\mb{W}}^k \hat{\mb{y}}^{k}_i + \hat{\mb{b}}^k), \forall k \in \{2, 3, \cdots, o\},\\
\hat{\mb{x}}_i &= \sigma(\hat{\mb{W}}^1 \hat{\mb{y}}^{1}_i + \hat{\mb{b}}^1).
\end{cases}$$

The objective of the auto-encoder model is to minimize the differences between the original feature vector $\mb{x}_i$ and the reconstructed feature vector $\hat{\mb{x}}_i$ of all the nodes in the network. Different from the traditional autoencoder model, to avoid trivial solutions, we propose to add a mask vector $\mb{c}_i$ in counting the introduced loss. In addition, to simplify the model parameter setting, we assume the user and post node embedding models will share the same set of parameters, and we will not differentiate the node types at this step. Formally, the embedding loss term can be represented as
$$\mathcal{L}_a(G) = \sum_{v_i \in \mathcal{V}} \left \| ({\mb{x}}_i - \hat{\mb{x}}_i) \odot \mb{c}_i \right\|_2^2.$$
Here, vector $\mb{c}_i$ is the weight vector corresponding to feature vector $\mb{x}_i$ of node $v_i$~\cite{zhang2017bl}. 
%Here, vector $\mb{c}_i$ is the weight vector corresponding to feature vector $\mb{x}_i$ of node $v_i$. Entries in vector $\mb{c}_i$ are filled with value $1$s except the entries corresponding to non-zero elements, which will be assigned with value $\gamma$ ($\gamma > 1$ denoting a larger weight to fit these features). The reason to add vector $\mb{c}_i$ is due to the sparsity of feature vector $\mb{x}_i$, in which majority of the entries are $0$s. If we simply set the reconstructed feature vector $\hat{\mb{x}}_i$ as $\mb{0}$, the introduced loss will still be minor. To avoid obtaining the trivial result, we give these non-zero feature a much higher weight $\gamma$ and try to preserve them in $\hat{\mb{x}}_i$. Here, we need to add a remark that ``simply discarding the entries corresponding zero values in the input vectors from the loss function'' will not work, since it will allow the model to decode their entries to any random values in $\hat{\mb{x}}_i$ on the other hand, which will not be what we want.

%-------------------------------------------------------------------------------------------------------------------------------------------------------------
%-------------------------------------------------------------------------------------------------------------------------------------------------------------
%-------------------------------------------------------------------------------------------------------------------------------------------------------------

\subsection{Collective Network Embedding Model}\label{subsec:embedding_model}

Different from the embedding problems studied for data instances which are independent of each other, the embedding process of nodes in online social networks are actually strongly correlated. Such correlations can be effectively quantified with the $n_{th}$-order \textit{diffusive network proximity} measure computed based on the heterogeneous network structure.

\subsubsection{Diffusive Network Proximity}

For the users who are close, like connected by friendship links or have replied to the same posts, they tend to be closer in the feature space. Such a closeness will ``constrain'' the distribution of their learned embedding feature vectors, in the feature space. It is similar for the post nodes as well. Based on the complex network connections in set $\mathcal{E}$, we can represent the connections among nodes as the \textit{adjacency matrix} $\mb{A} \in \{0, 1\}^{|\mathcal{V}| \times |\mathcal{V}|}$. Entry $A(i,j) = 1$ iff $(v_i, v_j) \in \mathcal{E}$ ($v_i, v_j \in \mathcal{V}$); otherwise, $A(i,j)$ will be filled with value $0$ instead. The adjacency matrix is also called the \textit{network local proximity} matrix in this paper.

Meanwhile, for the node pairs who are not connected by existing links, i.e., those corresponding to the $0$ entries in matrix $\mb{A}$, determining their closeness is a big challenging. So far, lots of network embedding models cannot handle these unconnected nodes well and will project them to random regions. Nowadays, some works propose the 2nd-order proximity for the closeness calculation based on 2-hop connections. However, these methods actually didn't solve the problem, as they still cannot figure out the closeness for the nodes which are not connected by either 1-hop or 2-hop connections. In this paper, we propose the ``diffusive proximity'' concept to help calculate the closeness of node pairs via literally $\infty$-order connections, namely the \textit{network global proximity}.
Based on the adjacency matrix $\mb{A}$, we introduce the normalized transition matrix $\mb{B}$, whose $(i_{th}, j_{th})$ entry denotes $B(i,j) = \frac{A(i,j)}{\sqrt{\sum_{i = 1}^{|\mathcal{V}|} A(i,j)} \sqrt{\sum_{j = 1}^{|\mathcal{V}|} A(i,j)}}$. Formally, matrix $\mb{B}$ can be formally represented as
$$\mb{B} = \mb{D}^{-\frac{1}{2}} \mb{A} (\mb{D}^\top)^{-\frac{1}{2}},$$
where $\mb{D}$ is the corresponding diagonal matrix of $\mb{A}$. For symmetric matrix $\mb{A}$, the normalized matrix $\mb{B}$ will still be symmetric, and information in matrix $\mb{B}$ denotes the normalized $1_{st}$-order local network proximity.

By multiplying the original adjacency matrix $\mb{A}$ with itself again, the resulting matrix contains the number of paths between node pairs via 2-hops. By following such an intuition, based on the normalized transition matrix, we can obtain the $2_{nd}$-order \textit{proximity matrix} $\mb{B}_2 = (\mb{B})^2$, as well as the $n_{th}$-order \textit{proximity matrix} $\mb{B}_n = (\mb{B})^n$. Meanwhile, different types of nodes and links have the different impact in steering the closeness among the nodes, which can be resolved by assigning the paths via different types of nodes with different weights. In this paper, to simplify the model settings, we will treat different node and link types equally. As $n$ increases, the proximity scores contained in $\mb{B}_n$ can capture broader network global information, which may also converge to a stable state. In this paper,  the resulting matrix $\mb{B}_{\bar{n}}$ will be used as the \textit{network global proximity matrix}. For a social network, it is easy to abtain the network global proximity matrix with a small $n$ due to the achievement of social science.
%Meanwhile, for some special types of networks or special structures, e.g., bi-partited networks, the convergence can actually never be obtained. It is hard to get rid of such structures form the network. Therefore, in this paper, instead of computing the stable state, we propose to increase $n$ to a feasibly large number $\bar{n}$, and the resulting matrix $\mb{B}_{\bar{n}}$ will be used as the \textit{network global proximity matrix}. For a social network, it is easy to abtain the network globa proximity matrix with a small $n$ due to the development of social science.

Formally, based on the learned embedding representations in $\{\mb{z}_i\}_{v_i \in \mathcal{V}}$, with the autoencoder model introduced in the previous subsection, the potential connection probability between node pair $(v_i, v_j)$ can be modeled as
$$p(v_i, v_j) = \frac{1}{1 + \exp(- \mb{z}_i^\top \cdot \mb{z}_j)}.$$
Here, the closer $\mb{z}_i$ and $\mb{z}_j$ are in the embedding feature space, the larger will the probability term $p(v_i, v_j)$ will be.
Meanwhile, by modeling the computed $n_{th}$-order \textit{network global proximity matrix} $\mb{B}_n$, for the node pair $(v_i, v_j)$, their connection probability can be effectively represented as value $\hat{p}(v_i, v_j) = B_n(i,j)$. In this paper, we propose to introduce a ``constraint'' on the node representations in the feature space based on the proximity matrix $\mb{B}_n$, based on the KL-divergence between distributions $p(\cdot, \cdot)$ and $\hat{p}(\cdot, \cdot)$ as follows:
$$\mathcal{L}_{n}(G) = - \sum_{v_i \in \mathcal{V}} \sum_{v_j \in \mathcal{V}, v_i \neq v_j} B_n(i,j) \log p(v_i, v_j).$$

\subsubsection{Collective Heterogeneous Social Network Embedding Objective Function}

According to the above descriptions, by adding the loss function introduced in embedding the heterogeneous social information together with the \textit{network proximity} preservation terms, we can represent the objective function for \textit{collective heterogeneous social network embedding} as 
$$\mathcal{L}_e(G) =  \mathcal{L}_a(G) + \sum_{n = 1}^{\bar{n}} \alpha_n \cdot \mathcal{L}_{n}(G) + \theta \cdot \mathcal{L}_{reg},$$
where $\mathcal{L}_{reg} = \sum_{k = 1}^o (\left\| \mb{W}^k \right\|_F^2 + \left\| \hat{\mb{W}}^k \right\|_F^2)$ denotes the regularization term of variables of the {\our} model and $\alpha_n$ is the weight of the loss term corresponding to $\mb{B}_n$. By solving the above function, we can learn the {\our} model, as well as obtaining the embedding feature vectors of all nodes in the network, which captures \textit{diverse social information} and \textit{network proximity} from the $1_{st}$ order to the $n_{th}$ order. 

\subsection{Task Oriented Network Embedding}

In the previous subsections, we have introduced the general network embedding model for heterogeneous social networks, which is very extensible and can effectively incorporate the external application objectives, e.g., \textit{network alignment}, \textit{community detection} and \textit{information diffusion}.

\subsubsection{Application Task 1: Network Alignment}\label{sec:task1}

The online social network alignment problem aims at inferring anchor links for shared users in multiple networks, which can be modeled as a binary-classification problem. Formally, Given a pair of aligned social networks $G^{(1)}$ and $G^{(2)}$, we can represent the sets of potential and partially observed anchor links between them as $\mathcal{L} = \mathcal{U}^{(1)} \times \mathcal{U}^{(2)}$ and $\mathcal{A}^{(1,2)} \subset \mathcal{L}$ respectively (where $\mathcal{U}^{(1)}$ and $\mathcal{U}^{(2)}$ denote the user sets in these two networks respectively). Based on the observed anchor link set $\mathcal{A}^{(1,2)}$, the network alignment task aims to learn a mapping $f:\mathcal{L} \rightarrow \mathcal{Y} $ to infer the labels of all the links in the set $\mathcal{L}$, where $\mathcal{Y}$ is the pre-deifined label space $\{0,1\}$. In other words, all the links in set $\mathcal{A}^{(1,2)}$ will be labeled as positive, while those in set $\mathcal{L}$ will be unlabeled instead, involving a mixture of both positive and negative instances. However, different from traditional classification problem, there exists an inherent \textit{one-to-one} cardinality constraint on the anchor links \cite{zhang2017link}. In this paper, we will follow the existing research works on network alignment \cite{zhang2018}, and adopt a linear model to infer the potential labels of anchor link instances based on their feature representations. In the case where the instances are not linearly separable, advanced methods like \textit{kernel tricks} \cite{elisseeff2002kernel} can be adopted to project the data instances into a high-dimensional feature space. By concatenating the learned embedding feature vectors of user nodes across networks, we can represent the features extracted for the anchor links as matrix $\mb{X} \in \mathbb{R}^{|\mathcal{L}| \times 2d}$ ($d$ denotes the learned embedding feature vector dimension). Meanwhile, the (potential) labels of all the anchor links in set $\mathcal{L}$ can be represented as vector $\mb{y} \in \{0, 1\}^{|\mathcal{L}|}$. To accomplish our constraint, we treat the network alignment task like a PU problem (\ie{ Positive and Unlabeled problem}). We can represent the introduced loss on the network alignment task by using the linear model as
$$ ||\mb{X} \mb{w} -\mb{y}||^2_2,$$
where $\mb{w}$ denotes the model feature weight variable vector. Now, considering \textit{one-to-one} cardinarity constraint in predicting anchor links, we model it as node degree constraint. Formally, we can define the network alignment task objective function to be
\begin{align*}
\mathcal{L}_{t}(G) = \min_{\mb{w}, \mb{y}} &\ \  \frac{1}{2}\cdot ||\mb{w}||_2^2 + \frac{k}{2} \cdot ||\mb{X}\mb{w}-\mb{y}||_2^2, \\ 
s.t.  ~~~~ &\mb{y} \in \{0,1\}^{|\mathcal{L}|}, \mbox{ } y_{i,j} = 1,\forall (u_i^{(1)},u_j^{(2)}) \in \mathcal{A}^{(1,2)},\\
&  0 \leq \sum_{u_i^{(1)} \in \mathcal{U}^{(1)}} y_{i,j} \leq 1 , \forall u_j^{(2)} \in \mathcal{U}^{(2)}, \\
& 0 \leq \sum_{u_j^{(2)} \in \mathcal{U}^{(2)}} y_{i,j} \leq 1 , \forall u_i^{(1)} \in \mathcal{U}^{(1)}, 
\end{align*}
where $||\mb{w}||_2^2$ denotes the regularization term on the model variable. Parameter $k$ is the scalar used to adjust the weight of the loss term. The objective function $\mathcal{L}_{t}(G)$ above is actually a NP-hard problem. Here, we will use an two-pharse algorithm proposed in \cite{zhang2017link} to get the approximated results.
\subsubsection{Application Task 2: Community Detection}
Users in online social networks can be divided into a set of communities, where users with frequent social interactions should belong to the same community. Based on user social connections, users' social interaction frequency can be effectively quantified as the social adjacency matrix $\mb{A}_u \in \mathbb{R}^{|\mathcal{U}| \times |\mathcal{U}|}$. Given the network $G$ with user set $\mathcal{U}$, let $\mathcal{C} = \{\mathcal{U}_1, \mathcal{U}_2, \cdots, \mathcal{U}_k\}$ be the $k$ disjoint social communities detected from the online social network $G$. The quality of the detected community can be measured with various metric, like \textit{normalized cut} \cite{SM00}.
%$$\mbox{N-Cut}(\mathcal{C}; \mb{A}_u) = \frac{1}{2} \sum_{j=1}^k \frac{S(\mathcal{U}_j, \mathcal{U} \setminus \mathcal{U}_j; \mb{A}_u)}{S(\mathcal{U}_j, \mathcal{U}; \mb{A}_u)},$$
%where $S(\mathcal{U}_j, \mathcal{U} \setminus \mathcal{U}_j; \mb{A}_u) = \sum_{u_l \in \mathcal{U}_j} \sum_{u_m \in \mathcal{U} \setminus \mathcal{U}_j} A_u(l,m)$ represents the number of connections between users in sets $\mathcal{U}_j$ and $\mathcal{U} \setminus \mathcal{U}_j$. Term $S(\mathcal{U}_j, \mathcal{U}; \mb{A}_u)$ denotes the connection volume (i.e., total number of connections) of users in $\mathcal{U}_j$.

Based on the embedding model introduced in Section~\ref{subsec:embedding_model}, let the latent representation vectors $\mb{z}_i \in \mathbb{R}^k$ denote the confidence scores for user $u_i$ belonging to the $k$ communities. Such community belonging indicator vectors can be organized as matrix $\mb{Z} = [\mb{z}_1^\top, \mb{z}_2^\top, \cdots, \mb{z}_{|\mathcal{U}|}^\top]^\top \in \mathbb{R}^{|\mathcal{U}| \times k}$. With matrix $\mb{Z}$, the normalized-cut based community detection objective function can be formally rewritten as
\vspace{-10pt}
$$\mbox{N-Cut}(\mathcal{C}; \mb{A}_u) = \frac{1}{2} \mbox{Tr}(\mb{Z}^\top \mb{L}_{\mb{A}_u} \mb{Z}),$$
where $\mb{L}_{\mb{A}_u} = \mb{D}_{\mb{A}_u} - \mb{A}_u$ denotes the Laplacian matrix corresponding to social adjacency matrix $\mb{A}_u$ and diagonal $\mb{D}_{\mb{A}_u}$ contains value $D_{\mb{A}_u}(i,i) = \sum_{j = 1}^{|\mathcal{U}|} A_u(i,j)$ on its diagonal. 
Furthermore, to avoid partitioning user nodes into multiple communities simultaneously, an orthonormal constraint is added to the indicator matrix $\mb{Z}$, i.e., $\mb{Z}^\top \mb{Z} = \mb{I}$. Such an orthonormal constraint renders the function extremely hard to solve, since the orthogonality constraints can lead to many local minimizers and, in particular, some of such orthonormal-constrained optimization problems in special forms are NP-hard. In this paper, we propose to relax the orthonormal constraint by replacing it with an orthonormal loss term $\left\| \mb{Z}^\top \mb{Z} - \mb{I} \right\|_F^2$ with a large weight instead. Therefore, the objective function for community detection application task can be formally represented as
\vspace{-10pt}
$$\mathcal{L}_t(G) = \frac{1}{2} \mbox{Tr}(\mb{Z}^\top \mb{L}_{\mb{A}_u} \mb{Z}) + \beta \cdot \left\| \mb{Z}^\top \mb{Z} - \mb{I} \right\|_F^2,$$
where $\beta$ denotes the weight (with a large value) of the loss term corresponding to the orthonormal constraint.

\subsubsection{Application Task 3: Information Diffusion}

Via the users' social interactions, information can diffuse from the initiators to other users in the network. For each topic $o_i \in \mathcal{O}$, its diffusion and infection trace can be outlined as the infected network $G_{o_i}$. Generally, for the users who have been activated by the same topics frequently, they tend to have closer preference and interest, which can be indicated from their learned embedding feature vectors in {\our}. Given a user pair $u_j$, $u_k$ in $G_{o_i}$, their relationship be categorized as the following four cases: (1) $u_j, u_k \in \mathcal{U}_{o_i} \land (u_j, u_k) \in \mathcal{E}_{o_i}$, (2) $u_j, u_k \in \mathcal{U}_{o_i} \land (u_j, u_k) \notin \mathcal{E}_{o_i}$, (3) $u_j \in \mathcal{U}_{o_i} \land u_k \notin \mathcal{U}_{o_i}$, and (4) $u_j \notin \mathcal{U}_{o_i} \land u_k \notin \mathcal{U}_{o_i}$. Among these $4$ cases, we can observe that case (1) indicates the strongest relation between $u_j$ and $u_k$, and then comes case (2), while case (3) denotes $u_j$ and $u_k$ have totally different interest in topic $o_i$, and case (4) shows no signal about their preference merely based on $G_{o_i}$. Let the embedding vector $\mb{z}_i$ denote the interest of user $u_i$ in the network, based on such an intuition, we introduce the following \textit{preference inequality} on embedding vectors:

\noindent \textbf{Definition 4} (Preference Inequality): Regarding topic $o_i$, given the user pairs (1) $(u_j, u_k) \in \mathcal{E}_{o_i}$, (2) $u_j, u_l \in \mathcal{U}_{o_i} \land (u_j, u_l) \notin \mathcal{E}_{o_i}$, (3) $u_j \in \mathcal{U}_{o_i}, u_m \notin \mathcal{U}_{o_i}$, and (4) $u_m \notin \mathcal{U}_{o_i}, u_n \notin \mathcal{U}_{o_i}$, we can represent the \textit{preference inequality} about the embedding feature vectors of users $u_j$, $u_k$, $u_l$, $u_m$ and $u_n$ as follows:
$$ \left\| \mb{z}_j - \mb{z}_k \right\|_2^2 \le \left\| \mb{z}_j - \mb{z}_l \right\|_2^2 \le \left\| \mb{z}_m - \mb{z}_n \right\|_2^2 \le \left\| \mb{z}_j - \mb{z}_m \right\|_2^2.$$

Furthermore, based on all the topics in $\mathcal{O}$, a set of \textit{preference inequality} equations can be defined, which will effectively constrain the relative distance of the learned embedding vectors. However, these inequality constraints may also lead to serious computation problems: (1) \textit{infeasible solution} as these constraints significantly shrink the feasible space and may result in no feasible solutions; (2) \textit{high computation cost} as these constraints are very challenging to preserve and will lead to very high computational costs. To overcome these challenges, we propose to relax the above \textit{preference inequality} equations and replace them with the following \textit{preference representation} objective function

\begin{multline}
 \mathcal{L}_{t}(G) = \sum_{o_i \in \mathcal{O}} \sum_{u_j, u_k \in \mathcal{U}} s^{o_i}_{j,k} \left\| \mb{z}_j - \mb{z}_k \right\|_2^2,  \\
 \mbox{where} \ \ 
s^{o_i}_{j,k} = \begin{cases}
\eta, &\mbox{ if } u_j, u_k \in \mathcal{U}_{o_i} \land (u_j, u_k) \in \mathcal{E}_{o_i},\\
1, &\mbox{ if } u_j, u_k \in \mathcal{U}_{o_i} \land (u_j, u_k) \notin \mathcal{E}_{o_i},\\
\delta, &\mbox{ if } u_j \in \mathcal{U}_{o_i} \land u_k \notin \mathcal{U}_{o_i},\\
0, &\mbox{ if } u_j \notin \mathcal{U}_{o_i} \land u_k \notin \mathcal{U}_{o_i},
\end{cases}
\end{multline}

\noindent Here, $s^{o_i}_{j,k}$ denotes the weight of the feature vector loss term based on $G_{o_i}$, and parameters $\eta > 1$ and $\delta < 0$. By integrating the objective function in {\our} together with the above \textit{preference representation} function, we will able to learn embedding applicable to information diffusion tasks specifically. For some other application tasks, we can also represent their requirements on the embedding results as the function $\mathcal{L}_{t}$, which will be incorporated in {\our} for model training.

\subsubsection{Task Oriented Network Embedding Objective Function}
For the task oriented network embedding, the objective function needs to consider both the network embedding loss term as well as the objectives from the external application tasks with a parameter $c \in [0, 1]$ balances between these two objectives. Formally, we can represent the joint objective function to be
$$\min c \cdot \mathcal{L}_{e}(G) + (1-c) \cdot \mathcal{L}_{t}(G).$$
To minimize the above objective function, we utilize Stochastic Gradient Descent (SGD). 
%For the task oriented network embedding, the objective function needs to consider both the network embedding loss term as well as the objectives from the external application tasks. Formally, we can represent the joint objective function to be
%$$\min c \cdot \mathcal{L}_{e}(G) + (1-c) \cdot \mathcal{L}_{t}(G),$$
%parameter $c \in [0, 1]$ balances between these two objectives. To minimize the above objective function, we can utilize Stochastic Gradient Descent (SGD). To be more specific, the training process involves multiple epochs. In each epoch, the training data is shuffled and a minibatch of the instances are sampled to update the parameters with SGD. Such a process continues until either convergence or the training epochs have been finished.

%Here, we need to add a remark. In the case when the external application tasks have the consistent requirements about the embedding vector as the embedding model itself, then the parameter $c$ will not affect the performance of the embedding model or the results of the application tasks based on the learned embedding feature vectors. However, for some other tasks, the task objective also needs to fit other application oriented historical data as well besides the network structure itself, and a careful tuning of $c$ will become necessary and crucial. In this paper, we propose to apply grid search, and select a good parameter $c$. More information about it is available in the following Section~\ref{sec:experiment}.

%-----------------------------------------------

\section{Experiments}\label{sec:experiment}

To test the effectiveness of the proposed model, extensive experiments have been done on a real-world heterogeneous social network dataset. In this section, we will first provide a brief description about the dataset used in the experiments, and then introduce the experimental settings, experimental results and parameter sensitivity analysis for the previous mentioned three tasks.

\subsection{Experimental Setting: Network Alignment Oriented Network Embedding}
We will introduce the experiment settings including the detailed setups,  comparison methods and evaluation metrics for network alignment oriented network embedding in this part. The dataset we use is two online heterogeneous social networks: Twitter and Foursquare. The detailed description is in the~\cite{KZY13}.
\subsubsection{Experiment Setup}
For the network alignment task,  we have $3,388$ anchor links between Twitter and Foursquare, which are treated as the positive anchor link instances. Meanwhile, for the remaining non-existing anchor links between networks, they are regarded as the negative anchor link instances instead. Because of the imbalance between positive set and negative set in real dataset (i.e., the negative set is far larger than the positive set), to make a detailed study in the performance of {\our} when facing imbalance data, we sample negative link instances from negative set according to a certain ratio regarding with the number of positive set, which is controlled by parameter $\lambda \in \{1, 2, \cdots, 10\}$. Furthermore, to eliminate the variance of each training process, we adopt the 10-fold cross validation to partition our dataset into the training set and testing set under different sample ratio. We take 9 folds out of 10 as training set and the rest is the testing set. For network alignment task, we will get the feature representation of each user first and then concatenate user features for anchor links according to the user pairs. After obtaining the combined feature representations, we feed it into the network alignment objective function introduced in this paper in order to predict label for potential links. In our proposed model {\our}, we filtered the users who do not have profile information since they cannot provide any useful information and set the parameter in diffusive network proximity $\bar{n} = 4$ since the users in networks are almost all connected to each other if the diffusive order is greater than $4$.

%-----------------------------------------------------------------------
\begin{table*}[t]
%	\vspace{-10pt}
\caption{Performance on Network Alignment Task of the comparison methods (parameter $\lambda$ changes in $\{1, 2, \cdots, 10\}$, steps of diffusive network proximity $i \in \{1,2,3,4\}$, $\alpha_i=1000, \theta=0.01, c = 0.001, k=1$).}
\label{tab:link_prediction_result}
\centering
{\tiny
\begin{tabular}{lrcccccccccc}
\toprule
\multicolumn{2}{l}{ }&\multicolumn{10}{c}{Negative/Positive Ratio $\lambda$}\\
\cmidrule{3-12}
metric &method &$1$ &$2$  &$3$   &$4$   &$5$   &$6$   &$7$   &$8$    &$9$  &$10$ \\
\midrule
\multirow{4}{*}{\rotatebox{90}{Accuracy}}
&{\our} &\textbf{0.903$\pm$0.012} &\textbf{0.878$\pm$0.012} &\textbf{0.862$\pm$0.010} &\textbf{0.856$\pm$0.008} &\textbf{0.854$\pm$0.008} &\textbf{0.860$\pm$0.007 }&\textbf{0.870$\pm$0.007} &\textbf{0.877$\pm$0.011} &\textbf{0.893$\pm$0.010 }&\textbf{0.903$\pm$0.006} \\
%\cmidrule{2-12}
&{\nodevec}&0.494$\pm$0.008  &0.659$\pm$0.015 &0.750$\pm$0.008 &0.800$\pm$0.008 &0.833$\pm$0.008  &0.857$\pm$0.004 &0.875$\pm$0.003 &0.889$\pm$0.004 &0.900$\pm$0.008 &0.909$\pm$0.005 \\
&{\autoencoder}&0.495$\pm$0.021 &0.665$\pm$0.012 &0.750$\pm$0.008 &0.800$\pm$0.008 &0.833$\pm$0.008 &0.857$\pm$0.004 &0.875$\pm$0.003 &0.889$\pm$0.004 &0.900$\pm$0.008 &0.909$\pm$0.005 \\
&{\deepwalk}&0.498$\pm$0.016 &0.666$\pm$0.012 &0.750$\pm$0.008 &0.800$\pm$0.008 &0.833$\pm$0.008 &0.857$\pm$0.004 &0.875$\pm$0.003 &0.889$\pm$0.004 &0.900$\pm$0.008 &0.909$\pm$0.005  \\
\cmidrule{1-12}

\multirow{4}{*}{\rotatebox{90}{F1}}
&{\our} &\textbf{0.910$\pm$0.013} &\textbf{0.840$\pm$0.018} &\textbf{0.774$\pm$0.012} &\textbf{0.719$\pm$0.014} &\textbf{0.674$\pm$0.017} &\textbf{0.641$\pm$0.014 }&\textbf{0.609$\pm$0.014} &\textbf{0.565$\pm$0.023} &\textbf{0.542$\pm$0.028} &\textbf{0.497$\pm$0.028} \\

%\cmidrule{2-12}
&{\nodevec} &0.489$\pm$0.017 &0.042$\pm$0.018 &0.000$\pm$0.000 &0.000$\pm$0.000 &0.000$\pm$0.000  &0.000$\pm$0.000  &0.000$\pm$0.000  &0.000$\pm$0.000  &0.000$\pm$0.000  &0.000$\pm$0.000  \\
&{\autoencoder} &0.498$\pm$0.027 &0.010$\pm$0.007 &0.002$\pm$0.004 &0.000$\pm$0.000  &0.000$\pm$0.000  &0.000$\pm$0.000  &0.000$\pm$0.000  &0.000$\pm$0.000  &0.000$\pm$0.000  &0.000$\pm$0.000   \\
&{\deepwalk} &0.496$\pm$0.031 &0.002$\pm$0.004 &0.000$\pm$0.000  &0.000$\pm$0.000  &0.000$\pm$0.000  &0.000$\pm$0.000  &0.000$\pm$0.000  &0.000$\pm$0.000  &0.000$\pm$0.000  &0.000$\pm$0.000  \\
\cmidrule{1-12}

\multirow{4}{*}{\rotatebox{90}{Precision}}
&{\our} &\textbf{0.849$\pm$0.021} &\textbf{0.745$\pm$0.022} &\textbf{0.655$\pm$0.015} &\textbf{0.589$\pm$0.016}&\textbf{0.537$\pm$0.020} &\textbf{0.505$\pm$0.014} &\textbf{0.488$\pm$0.014} &\textbf{0.468$\pm$0.033} &\textbf{0.477$\pm$0.041} &\textbf{0.472$\pm$0.032} \\
%\cmidrule{2-12}
&{\nodevec} &0.494$\pm$0.023 &0.317$\pm$0.113 &0.000$\pm$0.000 &0.000$\pm$0.000 &0.000$\pm$0.000 &0.000$\pm$0.000 &0.000$\pm$0.000 &0.000$\pm$0.000 &0.000$\pm$0.000 &0.000$\pm$0.000  \\
&{\autoencoder} &0.496$\pm$0.026 &0.303$\pm$0.223 &0.117$\pm$0.236 &0.000$\pm$0.000 &0.000$\pm$0.000 &0.000$\pm$0.000 &0.000$\pm$0.000 &0.000$\pm$0.000&0.000$\pm$0.000 &0.000$\pm$0.000 \\
&{\deepwalk}&0.500$\pm$0.023 &0.200$\pm$0.400 &0.000$\pm$0.000 &0.000$\pm$0.000 &0.000$\pm$0.000 &0.000$\pm$0.000 &0.000$\pm$0.000 &0.000$\pm$0.000 &0.000$\pm$0.000 &0.000$\pm$0.000 \\
\cmidrule{1-12}

\multirow{4}{*}{\rotatebox{90}{Recall}}
&{\our} &\textbf{0.980$\pm$0.006} &\textbf{0.963$\pm$0.012} &\textbf{0.948$\pm$0.009} &\textbf{0.923$\pm$0.016} &\textbf{0.905$\pm$0.018} &\textbf{0.875$\pm$0.014} &\textbf{0.810$\pm$0.037} &\textbf{0.716$\pm$0.017} &\textbf{0.630$\pm$0.032} &\textbf{0.526$\pm$0.032}  \\

%\cmidrule{2-12}

&{\nodevec} &0.485$\pm$0.026 &0.023$\pm$0.010 &0.000$\pm$0.000 &0.000$\pm$0.000 &0.000$\pm$0.000 &0.000$\pm$0.000 &0.000$\pm$0.000 &0.000$\pm$0.000 &0.000$\pm$0.000 &0.000$\pm$0.000\\
&{\autoencoder} &0.503$\pm$0.052 &0.005$\pm$0.004 &0.001$\pm$0.002 &0.000$\pm$0.000 &0.000$\pm$0.000 &0.000$\pm$0.000 &0.000$\pm$0.000 &0.000$\pm$0.000 &0.000$\pm$0.000 &0.000$\pm$0.000  \\
&{\deepwalk} &0.498$\pm$0.066 &0.001$\pm$0.002 &0.000$\pm$0.000 &0.000$\pm$0.000 &0.000$\pm$0.000 &0.000$\pm$0.000 &0.000$\pm$0.000 &0.000$\pm$0.000 &0.000$\pm$0.000 &0.000$\pm$0.000 \\
%\cmidrule{1-12}

\bottomrule

\end{tabular}
}
\vspace{-10pt}
%----------------------------------------------------------------------- 
\end{table*}

%\vspace*{-10pt}
\subsubsection{Comparison Methods}\label{subsec:comparison1}

This paper focuses on improving the existing network embedding model for the application oriented tasks, and the comparison methods used in this paper are mainly about the network embedding models, which are listed as follows:
\begin{itemize}
	
	\item {\our}: Framework {\our} is the general external application task oriented network embedding model proposed in this paper. The objective function of {\our} covers both the network embedding and external tasks, and the leaned embedding representation feature vector can effectively both the network structures and the application task objectives.
	
	\item \textit{Auto-encoder Model}: The {\autoencoder} model proposed in \cite{BLPL06} can project the instances into a low-dimensional feature space. In the experiments, we build the {\autoencoder} model merely based on the friendship link among users, and we also adjust the loss term for {\autoencoder} by weighting the non-zero features more with parameter $\gamma$ as introduced in Section~\ref{subsec:autoencoder}.
	
	\item \textit{Node2vec Model}: The {\nodevec} model \cite{GL16} adopts a flexible notion of a node's network neighborhood and design a biased random walk procedure to sample the neighbors. {\nodevec} can capture $1-k_{th}$-order of node proximity in homogeneous networks (based on users and their friendship connections).
	
	\item \textit{DeepWalk Model}: The {\deepwalk} model \cite{PAS14} extends the word2vec model \cite{MSCCD13} to the network embedding scenario. {\deepwalk} uses local information obtained from truncated random walks to learn latent based on social connections.
\end{itemize}
%\vspace{-10pt}

\subsubsection{Evaluation Metrics} To evaluate the performance of the learned embedding vectors, we use accuracy, precision, recall, and F1 as the evaluation metrics.
%\vspace{-10pt}

\subsection{Experimental Result: Network Alignment Oriented Network Embedding}
 In this part, we will show the experimental results of network alignment task. For {\our}, we set model parameters $k=1$, $c =0.001$, $\alpha_i = 1000$ for $i \in \{1,2,3,4\}$, and the regularization weight $\theta = 0.01$. Based on the parameter settings, we get the experiment results of network alignment task, which are shown in Table~\ref{tab:link_prediction_result}. Table~\ref{tab:link_prediction_result} can be divided into four parts and each part shows the result of one comparison method. A certain value (including the mean and standard deviation of 10-fold cross validation) in the table represents the result of the corresponding metrics under a certain negative/positive ratio and a certain comparison method. When taking a look at the table, it generally shows that our proposed method {\our} can achieve the best performance in the network alignment task for all four evaluation metrics. However, when the negative/positive ratio changing from 1 to 10, the performance of {\our} gradually gets worse. By contrast, we notice that the other three comparison methods cannot handle network alignment problem especially when the imbalance ratio get larger. The phenomenon suggests that the network structure provides less information to deal with the alignment task. The main reason is that these 3 baseline methods do not take considerations about the external application tasks in the embedding process. Such a detachment renders their embedding representations useless for the network alignment task. Whereas {\our} expresses the relatively strong robustness from experiment results. This is because we incorporate both the application task loss function and the one-to-one constraint in the objective function. Assisted by the application task, {\our} is able to learn very good embedding representations for the user nodes, which have shown to be effective for the network alignment task.

%We further analyze the parameter $c$ balancing the embedding loss term and the external application task objective function term. In Figure~\ref{fig:alignment_c}, by fixing the negative/positive ratio to be 1, we provide the performance of {\our} evaluated by Accuracy, F1, Precision and Recall by changing $c$ with values in $\{10^{-10}, 10^{-9}, \cdots, 10^{-1}\}$. According to the plot, we can see that when $c$ becomes larger, the overall trends of three metrics except Recall all go down. When $c$ comes to a certain point like $10^{-5}$, the performance of {\our} does not have a big change as $c$ further increases. However, for the Recall, its score curve goes up when $c$ grows larger. It indicates that with a larger $c$, more positive labels will be identified correctly, but the model will also mis-classify many negative instances to be positive by mistake.  

% \begin{figure}
%	\begin{minipage}[l]{0.5\textwidth}
%		\centering
%		\includegraphics[width=6cm]{./c_alignment.pdf}
%		\caption{Network Alignment Oriented Network Embedding Framework Parameter
%		 Analysis.}\label{fig:alignment_c}
%	 \vspace{-10pt}
%	\end{minipage}
%\end{figure}
\subsection{Experimental Setting: Community Detection Oriented Network Embedding}
In this part, we will introduce the experiment setting for the community detection oriented network embedding, including the detailed experiment setups, comparison methods, and evaluation metrics. 
%------------------------------------------
\begin{figure*}[htbp]
%	\vspace*{-10pt}
	\centering
	\subfigure[NDBI]{ \label{fig:community_detection_result_1}
		\begin{minipage}[l]{0.47\columnwidth}
			\centering
			\includegraphics[width=1.0\textwidth]{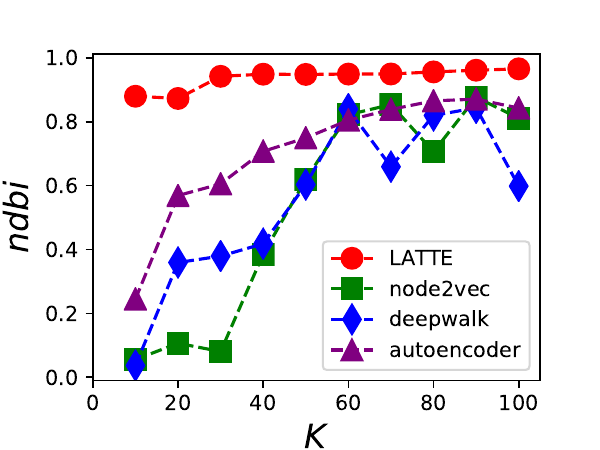}
		\end{minipage}
	}
	\subfigure[Silhouette]{\label{fig:community_detection_result_2}
		\begin{minipage}[l]{0.47\columnwidth}
			\centering
			\includegraphics[width=1.0\textwidth]{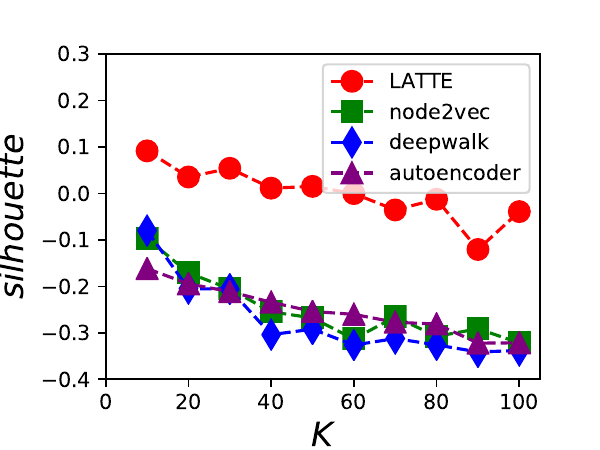}
		\end{minipage}
	}
	\subfigure[Density]{ \label{fig:community_detection_result_3}
		\begin{minipage}[l]{0.47\columnwidth}
			\centering
			\includegraphics[width=1.0\textwidth]{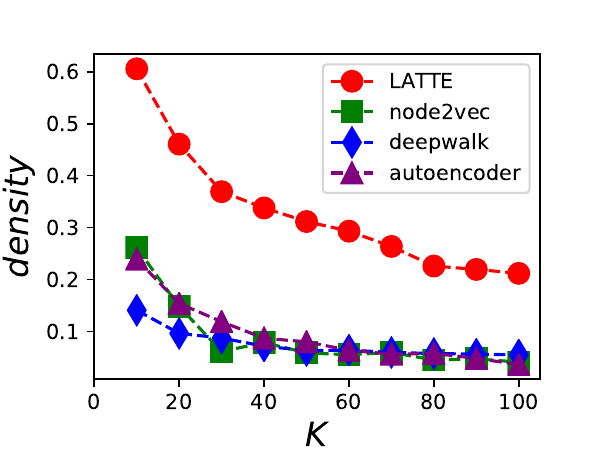}
		\end{minipage}
	}
	\subfigure[Entropy]{ \label{fig:community_detection_result_4}
		\begin{minipage}[l]{0.47\columnwidth}
			\centering
			\includegraphics[width=1.0\textwidth]{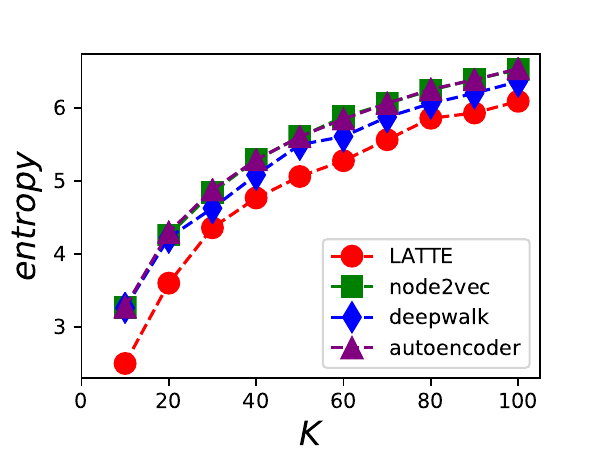}
		\end{minipage}
	}
	\vspace{-10pt}
	\caption{Community Detection Oriented Network Embedding Experimental Result.}\label{fig:community_detection_result}
	\vspace{-5pt}
\end{figure*}
%------------------------------------------

%\subsubsection{Experiment Setup}
%
%Based on the heterogeneous social networks, we can extract a set of initial feature representations of user and post nodes, which will be fed to the {\our} embedding model. Meanwhile, based on the social network structure, we can define the optimization function for the community detection task, where the orthogonal constraint on the latent embedding matrix will be relaxed and represented as a loss term in the objective function. By solving the problem, we will be able to learn the latent feature representation of the user nodes (as well as post nodes, which will not be used in the experiments). The network social community structure can be identified effectively by feeding the latent feature representations to the clustering algorithms, like KMeans. There is no good method to select the optimal community number, and we will try different community numbers $k \in \{10, 20, \cdots, 90, 100\}$.
%
%In the experiments, $5$ hidden layers are involved in framework {\our} (2 hidden layers in encoder step, 2 in decoder step, and 1 fusion hidden layer). The number neuron in these hidden layers are $256$, $128$, $k$, $128$ and $256$ respectively (i.e., the objective feature space dimension parameter $d = k$). The parameters $\alpha = 1.0$, $\beta = 1000.0$, $\theta = 0.1$, $\gamma = 1000.0$ and $0.001$ learning rate are used in the experiments.

\subsubsection{Comparison and Evaluation Metrics}
We will use the same methods introduced in the previous subsection~\ref{subsec:comparison1} as our comparison in community detection, where the application task incorporated in {\our} will be changed to \textit{community detection} instead. 
To evaluate the community structure outputted by different comparison methods, we will use $4$ other widely applied metrics \textit{normalized-dbi} \cite{DB79}, \textit{silhouette index} \cite{R87}, \textit{density} \cite{S07}, and \textit{entropy} \cite{NC07} in this paper. Metrics \textit{ndbi}, \textit{silhouette} will be computed based on the ``diffusive proximity'' scores $\mb{B}_{\bar{n}}$ at the ``stable'' state, \textit{density} counts the number of edges in each of the community, and \textit{entropy} measures the distribution of the community sizes.

%------------------------------------------
\begin{figure*}[t]
%	\vspace{-10pt}
	\centering
	\subfigure[NDBI]{ \label{fig:community_detection_parameter_analysis_1}
		\begin{minipage}[l]{0.47\columnwidth}
			\centering
			\includegraphics[width=1.0\textwidth]{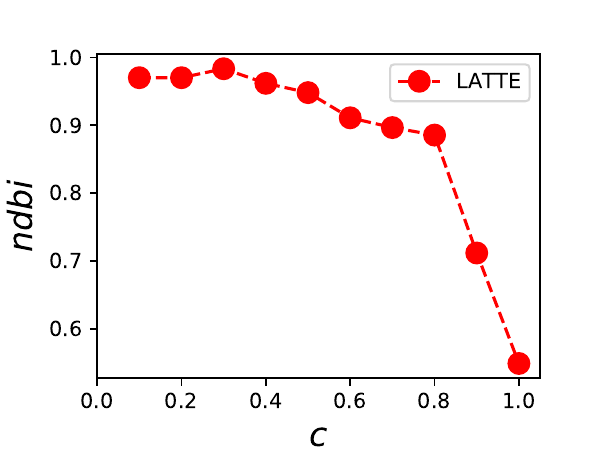}
		\end{minipage}
	}
	\subfigure[Silhouette]{\label{fig:community_detection_parameter_analysis_2}
		\begin{minipage}[l]{0.47\columnwidth}
			\centering
			\includegraphics[width=1.0\textwidth]{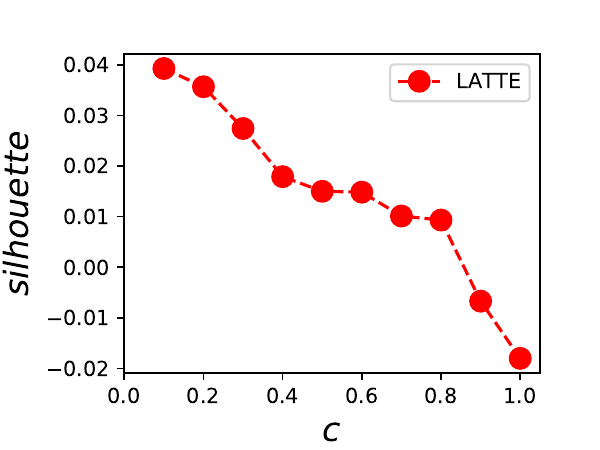}
		\end{minipage}
	}
	\subfigure[Density]{ \label{fig:community_detection_parameter_analysis_3}
		\begin{minipage}[l]{0.47\columnwidth}
			\centering
			\includegraphics[width=1.0\textwidth]{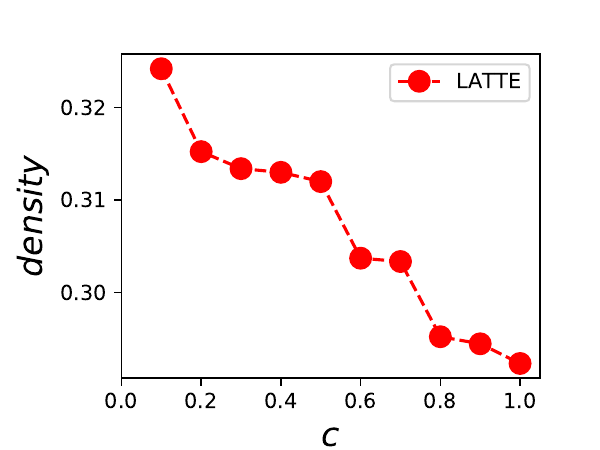}
		\end{minipage}
	}
	\subfigure[Entropy]{ \label{fig:community_detection_parameter_analysis_4}
		\begin{minipage}[l]{0.47\columnwidth}
			\centering
			\includegraphics[width=1.0\textwidth]{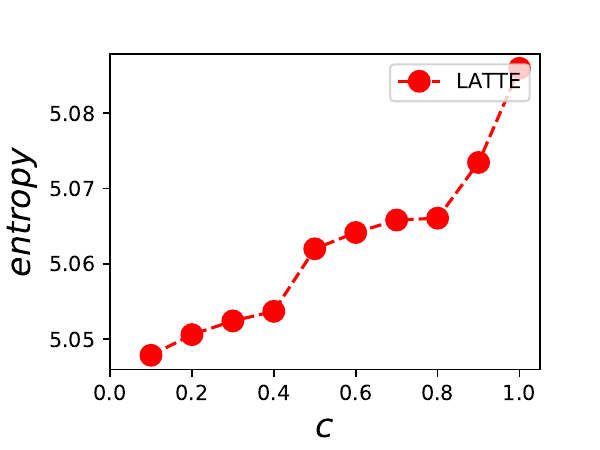}
		\end{minipage}
	}
\vspace{-8pt}
	\caption{Community Detection Oriented Network Embedding Framework Parameter Analysis.}\label{fig:community_detection_parameter}
	\vspace{-5pt}
\end{figure*}
%------------------------------------------

\subsection{Experimental Result: Community Detection Oriented Network Embedding} 
In Figure~\ref{fig:community_detection_result}, we show the experimental results of the community detection oriented network embedding task, evaluated by ndbi, silhouette index, density and entropy respectively. The x axis of the figures denotes $k$, i.e., the number of communities in the network, which changes in the range $\{10, 20, \cdots, 100\}$. Here, the parameter $c$ takes value $0.5$, denoting the embedding and community detection objectives have equal weights.

According to the results in Figure~\ref{fig:community_detection_result_1}, {\our} model incorporating the community detection objective in the framework learning can outperform the other pure network embedding models with great advantages. Metric ndbi (Normalized-DBI) effectively measures the number of links in communities against those between communities. According to the results, {\our} can achieve ndbi around 0.95 steadily for different numbers of communities, which denotes that the community detected by {\our} can generally partition closely connected user nodes into the same communities. The ndbi obtained by the other comparison methods are much lower than {\our}. For instance, when $k=10$ (i.e., we aim at partitioning the network into 10 communities), the ndbi scores obtained by {\autoencoder}, {\deepwalk} and {\nodevec} are all below 0.2, which is less than $\frac{1}{4}$ of the ndbi obtained by {\our}. In addition, the ndbi obtained by these methods varies a lot with different $k$ values, which indicates the unstableness of the results learned by these methods in community detection. Similar observations can be observed for the Silhouette metric in Figure~\ref{fig:community_detection_result_2}.

As the community number $k$ increases, the network will be partitioned into smaller communities, and more cross-community edges will be cut. According to the results in Figure~\ref{fig:community_detection_result_3}, the density of the community detection results achieved by all the methods will decrease as $k$ goes larger. Meanwhile, the density of the community detection obtained by {\our} is much larger and almost one time greater than those obtained by the other baseline methods. It indicates that {\our} will consider the edge cut loss in the embedding process, and the learned representation feature vectors can effective indicate the optimal community partition results of the network. In Figure~\ref{fig:community_detection_result_4}, we show the entropy obtained by all the comparison methods. Generally, entropy measures how balanced the network is partitioned in terms of community size, and balanced community structures (with close numbers of users) will achieve smaller entropy. According to the results, the community structures obtained by {\our} seems to be more balanced and reasonable compared with the community structures detected by the other methods. With detailed analysis of the community size in the results, the community detected by {\autoencoder}, {\deepwalk} and {\nodevec} contain some extremely small-sized and large-sized communities. 

In sum, according to the experimental results, incorporating the community detection objective in the network embedding learning process can effectively improve the effectiveness of the embedding results for the community detection task, which also demonstrates our claim at the beginning of this paper.

\subsection{Parameter Analysis: Community Detection Oriented Network Embedding}

In Figure~\ref{fig:community_detection_parameter}, we show the parameter sensitivity analysis about $c$ (i.e., the embedding loss weight) in the community detection oriented network embedding task with the evaluation metrics ndbi, silhouette, density and entropy respectively. In the analysis, we fix $k=50$ and change $c$ with values in $\{0.1, 0.2, \cdots, 0.9, 1.0\}$, where $1.0$ denotes the application task loss function will have weight $0$. According to the results, as $c$ value increases, the performance of {\our} will generally degrade steadily. The potential reason can be that,  as $c$ increase, the framework will aim optimizing the embedding component instead of the application task, and the learned embedding feature vectors can mainly reflect the embedding objective instead.

According to the results in Figure~\ref{fig:community_detection_parameter}, we can observe that when $c=1.0$, the performance of {\our} can still outperform the other baseline methods (with $k=50$) in Figure~\ref{fig:community_detection_result} respectively, except {\autoencoder} with metric ndbi. It shows that utilizing the heterogeneous information for the network representation learning in {\our} can helpfully capture better social community structure than the other network embedding methods.

\subsection{Experimental Setting: Information Diffusion Oriented Network Embedding}

In this part, we will introduce the experimental setting for the information diffusion oriented network embedding, including the detailed experiment setups, comparison methods, and evaluation metrics respectively. 
In the experiments, $5$ hidden layers are involved in framework {\our} (2 hidden layers in encoder step, 2 in decoder step, and 1 fusion hidden layer). The number neuron in these hidden layers are $256$, $128$, $64$, $128$ and $256$ respectively (i.e., the objective feature space dimension parameter $d = 64$). The parameters $\alpha = 1.0$, $\theta = 0.1$, $\eta = 10.0$, $\delta = -1.0$, $\gamma = 1000.0$ and $0.001$ learning rate are used in the experiments.

\subsubsection{Comparison Methods and Evaluation Metrics}

The comparison methods used in the information diffusion oriented network embedding task are generally identical to those introduced in Section~\ref{subsec:comparison1}, except the application objective function in {\our} will be changed to that of the information diffusion task instead. Based on the computed potential diffusion link probability, we can compute the AUC and Precision@100 by comparing them with the ground truth labels of the diffusion links in the test set. Furthermore, we also adopted a threshold $t \in [0, 1]$ to partition these prediction results into two bins, namely the positive set and negative set. For the diffusion link probability greater than $t$, we will assign them with a positive label; while for those with probability smaller than $t$, they will be assigned with a negative label instead. Various classification based evaluation metrics, including F1, Precision, Recall, Accuracy, will be used here to evaluate the quality of the predicted labels in the experiments.

\subsection{Experimental Result and Parameter Analysis: Information Diffusion Oriented Network Embedding} 

\begin{figure}[t]
	\centering
		\subfigure[AUC]{ \label{fig:information_diffusion_result_1}
			\begin{minipage}[l]{0.45\columnwidth}
				\centering
				\includegraphics[width=0.8\textwidth]{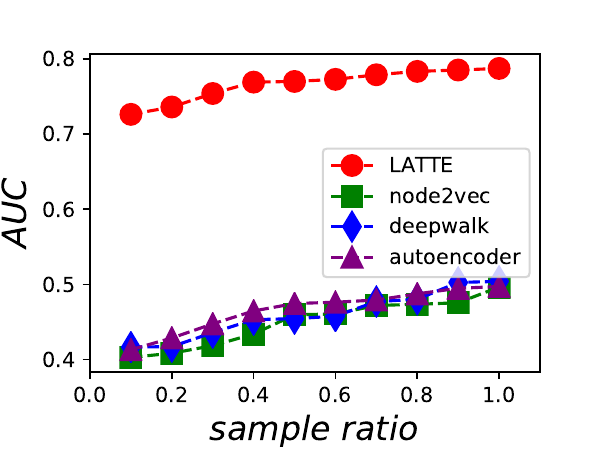}
			\end{minipage}
		}
		\subfigure[Prec@K]{\label{fig:information_diffusion_result_2}
			\begin{minipage}[l]{0.45\columnwidth}
				\centering
				\includegraphics[width=0.8\textwidth]{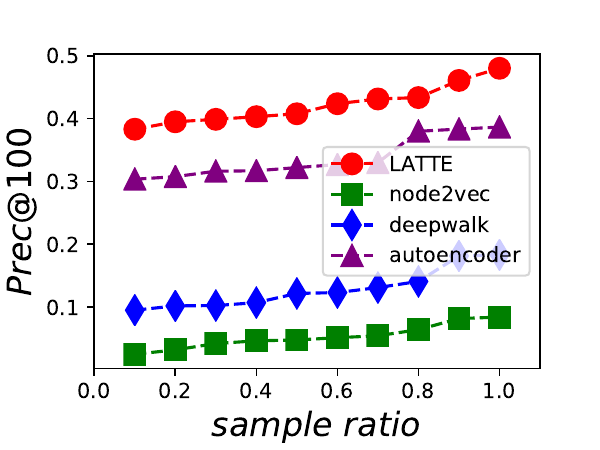}
			\end{minipage}
		}
	\vspace{-5pt}
		\caption{Result of Information Diffusion Task  Oriented Network Embedding Framework.}\label{fig:information_diffusion_result}
	\vspace{-10pt}
\end{figure}

In Figure~\ref{fig:information_diffusion_result}, we show the experimental results obtained by the comparison methods in the information diffusion oriented network embedding task, evaluated by AUC and Precision@100. Besides embedding the networks structure and social information, the objective studied here will also cover the modeling of information diffusion process in the networks. Here, to denote different ratios of labeled diffusion links, we change the sample ratio with values in $\{0.1, 0.2, \cdots, 0.9, 1.0\}$, and fix the weight $c$ with value $0.5$.

According to the performance of the comparison methods, the results obtained by methods {\deepwalk}, {\autoencoder} and {\nodevec} don't change with $c$. The main reason is that these methods don't use the information diffusion information in the model building, and the change the sample ration will not affect the performance of these methods. According to Figures~\ref{fig:information_diffusion_result_1} and \ref{fig:information_diffusion_result_2}, as the sample ratio increases, the AUC and Precision@100 achieved by {\our} will increase. The main reason is that with larger sample ratios, more positive diffusion links will be available in the training set and the learned model as well as the embedding feature vectors will be able to capture the patterns about these positive diffusion links. It will lead to better prediction performance on the testing set.

\begin{figure}[t]
	\centering
	\subfigure[AUC]{ \label{fig:information_diffusion_parameter_analysis_1}
		\begin{minipage}[l]{0.45\columnwidth}
			\centering
			\includegraphics[width=0.8\textwidth]{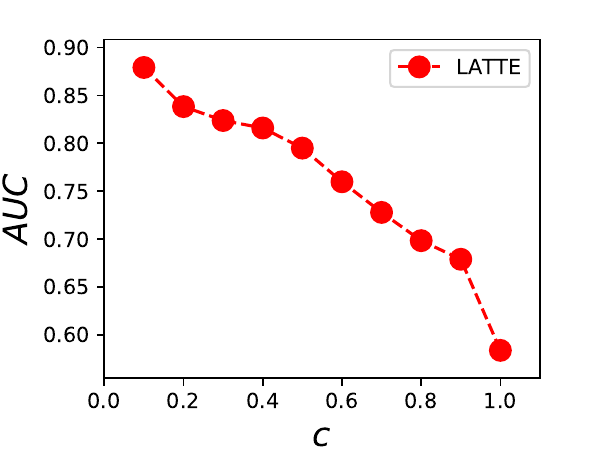}
		\end{minipage}
	}
	\subfigure[Precision@K]{\label{fig:information_diffusion_parameter_analysis_2}
		\begin{minipage}[l]{0.45\columnwidth}
			\centering
			\includegraphics[width=0.8\textwidth]{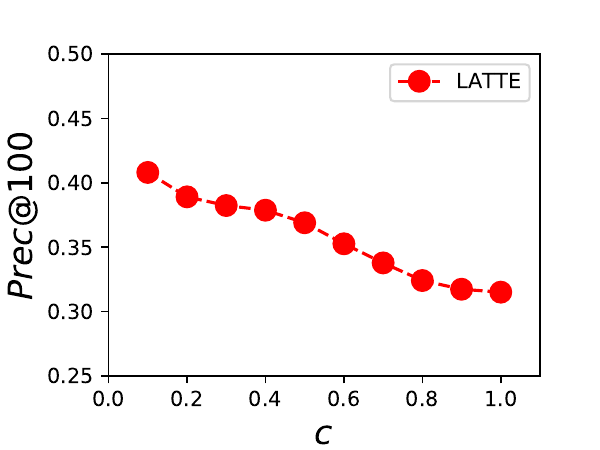}
		\end{minipage}
	}
\vspace{-5pt}
	\caption{Information Diffusion Oriented Network Embedding Framework Parameter Analysis.}\label{fig:information_diffusion_parameter}
	\vspace{-10pt}
\end{figure}

In Figure~\ref{fig:information_diffusion_parameter}, we provide the parameter sensitivity analysis about the weight $c$ in the framework. Similar observations can be found in the results, as $c$ increase, the AUC and Precision@100 score obtained by {\our} will go down. Slightly different from the community detection task, as $c$ goes to $1.0$, the performance metric scores of {\our} will be slightly lower than those obtained by {\autoencoder}. It indicates that incorporating the heterogeneous network structure and social information in the embedding process may not be very useful for the information diffusion task, and non-relevant information can be slightly misleading for inferring the potential future diffusion links among users in the networks.

%------------------------------------------
%\begin{figure}[t]
%	\centering
%
%		\subfigure[AUC]{ \label{fig:information_diffusion_result_1}
%			\begin{minipage}[l]{0.45\columnwidth}
%				\centering
%				\includegraphics[width=1.0\textwidth]{./auc.pdf}
%			\end{minipage}
%		}
%		\subfigure[Prec@K]{\label{fig:information_diffusion_result_2}
%			\begin{minipage}[l]{0.45\columnwidth}
%				\centering
%				\includegraphics[width=1.0\textwidth]{./prec100.pdf}
%			\end{minipage}
%		}
%		\caption{Information Diffusion Oriented Network Embedding Experimental Result.}\label{fig:information_diffusion_result}
%	
%\end{figure}
%%------------------------------------------

%-----------------------------------------------
\section{Related Work} \label{sec:related_work}

%---- embedding problems ----
% In graphs, the relation can be treated as a translation of the entities, and many translation based embedding models have been proposed, like TransE \cite{BUGWY13}, TransH \cite{WZFC14} and TransR \cite{LLSLZ15}.
\noindent \textbf{Network Representation Learning}: Network representation learning has become a very hot research problem recently, which can project a graph-structured data to the feature vector representations. In recent years, many network embedding works based on random walk model and deep learning models have been introduced, like Deepwalk \cite{PAS14}, node2vec \cite{GL16}, AutoNE \cite{tu2019autone} and CAN \cite{meng2019co}. Perozzi et al. extends the word2vec model \cite{MSCCD13} to the network scenario and introduce the Deepwalk algorithm. Chang et al. \cite{CHTQAH15} learn the embedding of networks involving text and image information. Chen et al. \cite{CS16} introduce a task guided embedding model to learn the representations for the author identification problem. Most of these embedding models are proposed for homogeneous networks, and assume the learned feature vectors can be applicable to all external tasks. These existing methods will suffer from great problems in the real-world applications mainly due to the inconsistency between specific task objectives against the embedding objective.

%	\begin{minipage}[b]{0.45\columnwidth}
%	\flushright
%	\subfigure[AUC]{ \label{fig:information_diffusion_parameter_analysis_1}
%		%		\begin{minipage}[l]{0.45\columnwidth}
%		%			\centering
%		\includegraphics[width=0.8\textwidth]{./c_auc.pdf}
%		%		\end{minipage}
%	}
%	\subfigure[Precision@K]{\label{fig:information_diffusion_parameter_analysis_2}
%		%		\begin{minipage}[l]{0.45\columnwidth}
%		%			\centering
%		\includegraphics[width=0.8\textwidth]{./c_prec.pdf}
%		%		\end{minipage}
%	}
%	\caption{Parameter Analysis of Information Diffusion Task.}\label{fig:information_diffusion_parameter}
%	%		\vspace{-10pt}
%\end{minipage}

%------------------------------------------
%\begin{figure}[t]
%	\centering
%	\subfigure[AUC]{ \label{fig:information_diffusion_parameter_analysis_1}
%		\begin{minipage}[l]{0.45\columnwidth}
%			\centering
%			\includegraphics[width=0.8\textwidth]{./c_auc.pdf}
%		\end{minipage}
%	}
%	\subfigure[Precision@K]{\label{fig:information_diffusion_parameter_analysis_2}
%		\begin{minipage}[l]{0.45\columnwidth}
%			\centering
%			\includegraphics[width=0.8\textwidth]{./c_prec.pdf}
%		\end{minipage}
%	}
%	\caption{Information Diffusion Oriented Network Embedding Framework Parameter Analysis.}\label{fig:information_diffusion_parameter}
%	\vspace{-10pt}
%\end{figure}
%------------------------------------------

\noindent \textbf{Network Alignment}. Network alignment problem is an important research problem, which has been studied in various areas, e.g., protein-protein-interaction network alignment in bioinformatics \cite{KBS08,LLBSB09,SXB07}, chemical compound matching in chemistry \cite{SHL08},  ontology alignment web semantics \cite{DMDH04}, graph matching in combinatorial mathematics \cite{MH14}, figure matching and merging in computer vision \cite{CFSV04,BGGSW09}, and online social network matching and alignment \cite{KZY13,ZYZ14,ZY15}. Recently, lots of works have been done for social network alignment specifically, IONE maintains the both local and higher-order structural information~\cite{liu2019structural}.  ActiveIter utilizes the meta diagram in a active learning framenwork~\cite{ren2019meta}. But None of these can be extensible for other tasks.

\noindent \textbf{Clustering and Community Detection}. Clustering is a very broad research area, which includes various types of clustering tasks, like consensus clustering \cite{LBRFFP13,LD13}, multi-view/relational clustering \cite{BS04,CNH13,YHY07}, co-training based clustering \cite{KD11}. In recent years, clustering based community detection in online social networks is very popular, where a comprehensive survey is available in \cite{MV13}. Several different techniques have been proposed to optimize certain community metrics, e.g., modularity \cite{NG04} or normalized cut \cite{SM00}. A detailed tutorial on spectral clustering has been given by Luxburg in \cite{L07}. In this paper, we propose to do community detection by incorporating it in the embedding task, where the learned embedding feature vectors can capture not only network structure but also the community structure at the same time.

\noindent \textbf{Influence Maximization and Information Diffusion}. Influence maximization problem as a popular research topic was first proposed by Domingos et al. \cite{DR01}. It was first formulated as an optimization problem in \cite{KKT03}, where Kempe et al. proposed two basic stochastic influence propagation models, the \textit{independent cascade (IC) model} and \textit{linear threshold (LT) model}.  Since then, a considerable amount of works on speeding up the seed selection algorithms \cite{chen2010scalable2} are introduced, including the scalable CELF model \cite{LKGFVG07} and heuristic based algorithms for IC \cite{CWY09} and LT \cite{CWW10} models. Information diffusion on heterogeneous and multi-relational networks has become an increasingly important topic in recent years \cite{SHYYW11,liu2010mining}. However, all these existing works mainly focus on constructing the models that fit the information diffusion process, but fail to learn users' topic preference representations as well as the diffusion patterns representations, which will be studied in this paper using the embedding approach.
%-----------------------------------------------
\section{Conclusion}\label{sec:conclusion}

In this paper, we have studied the ``\textit{application oriented network embedding}'' problem, which aims at learning the heterogeneous network embeddings subject to specific application requirements. To address the problem, we introduce a novel application oriented heterogeneous network embedding model, namely {\our}. The node closeness can be effectively measured with the novel ``diffusive proximity'' concept in {\our} based on $1-n_{th}$-order of node proximity. By extending the autoencoder model, the embedding results learned by {\our} can both preserve the heterogeneous network structure as well as incorporating the external application objectives effectively. Extensive experiments done on a real-world heterogeneous networked dataset have demonstrated the effectiveness and advantages of {\our} over other existing network embedding models in application tasks, like network alignment, community detection and information diffusion.
%-----------------------------------------------
\label{sec:ack}
\section{Acknowledgement}
This work is partially supported by NSF through grant IIS-1763365 and by FSU.

%-----------------------------------------------
\balance
\bibliographystyle{plain}
\bibliography{reference}

\end{document}